\documentstyle[prd,preprint,tighten,aps,eqsecnum,floats,%
amssymb,amsfonts,newlfont,graphicx]{revtex}
\begin{document}
\draft
\preprint{
\begin{tabular}{r}
   KIAS-P99011
\\ DFTT 08/99
\\ arXiv:hep-ph/9902462
\end{tabular}
}
\title{Finally neutrino has mass}
\author{S.M. Bilenky}
\address{
Joint Institute for Nuclear Research, Dubna, Russia
\\
and
\\
School of Physics,
Korea Institute for Advanced Study,
Seoul 130-012, Korea
}
\author{C. Giunti}
\address{
INFN, Sez. di Torino, and Dip. di Fisica Teorica,
Univ. di Torino,
I--10125 Torino, Italy
\\
and
\\
School of Physics,
Korea Institute for Advanced Study,
Seoul 130-012, Korea
}
\author{C.W. Kim}
\address{
School of Physics,
Korea Institute for Advanced Study,
Seoul 130-012, Korea
\\
and
\\
Dept. of Physics $\&$ Astronomy,
The Johns Hopkins University,
Baltimore, MD 21218, USA
}
\maketitle
\begin{abstract}
The present status of the problem of neutrino mass, mixing and neutrino
oscillations is briefly summarized.
The evidence for oscillations of atmospheric neutrinos found recently in the
Super-Kamiokande
experiment is discussed.
Indications in favor of neutrino oscillations obtained in
solar neutrino experiments and in the accelerator LSND experiment
are also considered.
Implications of existing neutrino oscillation data for
neutrino masses and mixing are discussed.
\end{abstract}


\section{Introduction}
\label{Introduction}

The Glashow-Weinberg-Salam \cite{SM}
theory of electroweak interactions,
which combined with the Quantum Chromo-Dynamics (QCD)
is now called the Standard Model (SM),
is one of the greatest achievements of  particle physics
in the 20$^{\mathrm{th}}$ century.
Among others,
the Glashow-Weinberg-Salam theory allowed to
predict successfully the existence of
charmed-particles \cite{GIM70},
of the $b$ and $t$ quarks,
a new class of weak interactions
(neutral currents),
the existence of the vector $W^{\pm}$ and $Z^0$ bosons and their masses. 
All the predictions of the SM have been confirmed by
numerous experiments; the theory
describes beautifully all the existing experimental data in the whole energy
range available at present
(except for the indications in favor of neutrino oscillations
that we will discuss in the following). 

However, the prevailing general consensus is that the SM cannot be
the final theory of elementary particles. 
The SM is a
theory of weak, electromagnetic and strong interactions with
the exception of gravity.
In this theory, more than 20 arbitrary fundamental parameters
(masses of quarks and leptons,
coupling constants, mixing angles, etc.)
still remain to be explained.
Also, there are several conceptual problems;
to name only two,
the lack of any explanation of why in nature
there exist three generations of quarks and
leptons that differ only in masses
and
the hierarchy problem,
connected with
the radiative corrections to
the mass of the Higgs boson.

Major efforts at the moment 
are directed towards the
search for a theory of elementary particles that could generalize the SM
and would solve the problems mentioned above.
In the past, many such models have been proposed.
They include, among others, Grand Unified models,
Supersymmetric models, Superstring models, composite
models
\cite{art:Ellis:1998eh}.
In many experiments, possible
effects of
physics
beyond the Standard Model have been and will be  searched for.
In the frontier of
accelerator high energy experiments, one of the major goals is to find,
as a signature of new physics, supersymmetric particles or some
unexpected behavior of the standard Higgs boson.
Among accelerator and non-accelerator
physics experiments,
one of the most popular searches have been
that of neutrino masses via neutrino oscillations,
for most theories beyond the SM predict non-zero neutrino masses,
and that of proton decay.
At present
some evidence on new
physics beyond the SM has been found
only in neutrino oscillation experiments.

If neutrinos are  massive and mixed,
the states of \textit{flavor neutrinos}  $\nu_e$, 
$\nu_{\mu}$, $\nu_{\tau}$
are mixed coherent superpositions of the states of neutrinos with definite
mass. In this case, neutrinos 
produced via weak interactions
experience \textit{neutrino oscillations},
which are
periodical transitions among different flavor neutrinos.
Such effects appear to have been observed in several neutrino
experiments.  

An impressive evidence for the disappearance of
atmospheric $\nu_\mu$'s 
has been presented by the
Super-Kamiokande collaboration
\cite{SK-atm-98,Scholberg-Venice-99,%
Kajita-Ringberg-99,Learned-JHU-99,Nakahata-TAUP-99}.
Similar indications in favor of
neutrino oscillations have also been obtained in the
Kamiokande \cite{Kam-atm},
IMB \cite{IMB},
Soudan 2 \cite{Soudan2}
and
MACRO \cite{MACRO}
atmospheric neutrino experiments.
All the existing data of solar
neutrino experiments
(Homestake \cite{Homestake},
Kamiokande \cite{Kam-sun},
GALLEX \cite{GALLEX},
SAGE \cite{SAGE},
Super-Kamiokande \cite{SK-sun,Nakahata-TAUP-99})
can naturally be explained by neutrino mass
and mixing.
Finally, some indication in favor of 
$\bar\nu_\mu\to\bar\nu_e$ and $\nu_\mu\to\nu_e$
oscillations has been found in the accelerator LSND experiment
\cite{LSND-96-PRL,LSND-98-PRL}.

These data constitute
\textit{the first observation of processes in which lepton numbers
are not conserved}.
It is a general belief that such phenomena are due to
physics beyond the SM
\cite{Wilczek-nu98}.

The purpose of this article is to review
the theory and phenomenology of neutrino masses,
neutrino mixing and the salient features of neutrino
oscillations.
In Section~\ref{Two-component neutrino, Standard Model and lepton numbers},
a short discussion of the
theory of two-component neutrinos,
the Standard Model and the law of conservation of lepton numbers will be
given.
In Section~\ref{Neutrino mass and mixings},
we will consider
the problem of neutrino mass and different possibilities of neutrino mixing.
Neutrino oscillations will be discussed in
Section~\ref{Neutrino oscillations}
and the experimental
data will be briefly presented and discussed in
Section~\ref{Status of neutrino oscillations}.
Conclusions are drawn in Section~\ref{Conclusions}.

\section{Two-component neutrino, Standard Model and lepton numbers}
\label{Two-component neutrino, Standard Model and lepton numbers}

In 1957,
soon after the discovery of parity violation in weak interactions
(Wu \textit{et al.} \cite{Wu57})
Landau \cite{Landau57},
Lee and Yang \cite{Lee-Yang57}
and
Salam \cite{Salam57}
proposed the theory of
two-component massless neutrinos.

Starting with the field $\nu(x)$ of the neutrino with mass $ m $,
which satisfies the Dirac equation
\begin{equation}
( i \gamma^{\alpha} \partial_{\alpha} - m ) \ \nu = 0
\,,
\label{Dirac}
\end{equation}
let us introduce the left-handed $\nu_L$ and right-handed $\nu_R$ components
of the neutrino field, respectively, as
\begin{equation}
\nu_{L,R} = \frac{1\mp\gamma_5}{2} \ \nu
\label{LR}
\,.
\end{equation}
From (\ref{Dirac}),
for $\nu_{L,R}$ we have a set of coupled equations,
\begin{eqnarray}
i \gamma^\alpha \partial _\alpha \nu_L - m \nu_R = 0
\,,
\nonumber
\\
i \gamma^\alpha \partial _\alpha \nu_R - m \nu_L = 0
\,.
\end{eqnarray}
If the neutrino mass is zero, the equations for  $\nu_L$ and
$\nu_R$ are decoupled,
\begin{equation}
i \gamma^{\alpha} \partial_{\alpha} \nu_{L} = 0
\,,
\qquad
i \gamma^{\alpha} \partial_{\alpha} \nu_{R} = 0
\,,
\label{Weyl}
\end{equation}
and 
the left-handed component $\nu_L$ (or right-handed component $\nu_R$) can
be chosen as the neutrino field.
This was the choice of the authors of the two-component neutrino theory.

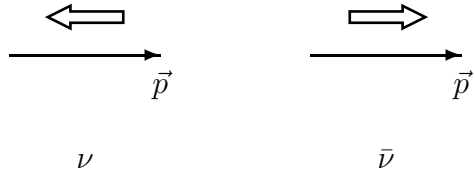
\begin{figure}[t]
\begin{center}
\setlength{\unitlength}{1cm}
\begin{picture}(6,2.5)
\thicklines
\put(0,1.5){\vector(1,0){2}}
\put(2,1.3){\makebox(0,0)[t]{$\vec{p}$}}
\put(1,0){\makebox(0,0)[b]{$\nu$}}
\put(4,1.5){\vector(1,0){2}}
\put(6,1.3){\makebox(0,0)[t]{$\vec{p}$}}
\put(5,0){\makebox(0,0)[b]{$\bar\nu$}}
\put(0,0){
 }
\end{picture}
\end{center}
\caption{ \label{Helicities}
Helicities of two component neutrino and antineutrinos
described by the field $\nu_L$.
The direction of momentum (thin arrows) and spin (thick arrows) are shown.}
\end{figure}

If the neutrino field is $\nu_L$,
a neutrino with definite momentum
has negative helicity
(projection of spin on the direction of the momentum) 
and an antineutrino has positive helicity (see Fig.~\ref{Helicities}).
If the neutrino field is $\nu_R$, a neutrino has positive helicity
and an antineutrino has negative helicity.

The two-component neutrino theory was confirmed in the
famous Goldhaber \textit{et al.} experiment \cite{Goldhaber58} (1958),
in which the
helicity of neutrino was determined from the measurement of circular 
polarization of
the photon in the process
\begin{eqnarray}
e^- + \mathrm{Eu} \to \nu_e + \null & \mathrm{Sm}^* &
\nonumber
\\
& \downarrow &
\nonumber
\\
& \mathrm{Sm} & \null + \gamma
\,.
\end{eqnarray} 
The measured circular polarization was consistent with a
\textit{neutrino with negative helicity}.

The phenomenological $V-A$ theory of weak interactions by
Feynman and Gell-Mann \cite{Feynman-Gell-Mann58}
and
Sudarshan and Marshak \cite{Sudarshan-Marshak58}
was based on the assumption that only the
\textit{left-handed components of all fields} are involved
in the Hamiltonian of weak interactions.
Assuming also the universality of the weak interactions,
Feynman and Gell-Mann in 1958  proposed a Hamiltonian 
that is a product of two currents
\begin{equation}
\mathcal{H}_I^{\mathrm{CC}}
=
\frac{G_{\mathrm{F}}}{\sqrt{2}} \ j^{\mathrm{CC}\alpha} \ {j^{\mathrm{CC}}_{\alpha}}^\dagger
\,,
\label{FermiCC}
\end{equation}
which was successful in describing all the existing  weak  interaction
data. 
In Eq.(\ref{FermiCC}),
$j_{\alpha}^{\mathrm{CC}}$ is the charged weak current (see later) and $G_{\mathrm{F}}$ is the
Fermi constant.

The SM is 
based on a spontaneously broken
$SU(2)\times U(1)$ local gauge group with the left-handed doublets
\begin{equation}
\left(\begin{array}{c} \nu_{e} \\ e \end{array} \right)_L
\,,
\qquad
\left(\begin{array}{c} \nu_{\mu} \\ \mu \end{array} \right)_L
\,,
\qquad
\left(\begin{array}{c} \nu_{\tau} \\ \tau \end{array} \right)_L
\,,
\label{left-handed doublets}
\end{equation}
and the right-handed singlets $e_R$, $\mu_R$ and $\tau_R$
(we are interested here only in the lepton part of the SM).

The interactions of leptons and vector bosons in the SM
has three parts:

\begin{enumerate}

\item
The Hamiltonian of electromagnetic interaction
\begin{equation}
\mathcal{H}_I^{\mathrm{em}} = e \ j^{\mathrm{em}}_{\alpha} \ A^{\alpha}
\,,
\label{Hem}
\end{equation}
where $A_\alpha$ is the electromagnetic field,
$e$ is the electric charge and
\begin{equation}
j^{\mathrm{em}}_{\alpha}
=
\bar{e} \gamma_{\alpha} e
+
\bar{\mu} \gamma_{\alpha} \mu
+
\bar{\tau} \gamma_{\alpha} \tau
\end{equation}
is the electromagnetic current;

\item
The Hamiltonian of charged current (CC) weak interactions
\begin{equation}
\mathcal{H}_I^{\mathrm{CC}}
=
\frac{g}{2\sqrt{2}} \ j^{\mathrm{CC}}_\alpha \ W^{\alpha} + \mathrm{h.c.}
\,,
\label{HCC}
\end{equation}
where $g$ is an interaction constant,
$W^\alpha $ is the field of charged vector bosons,
and
\begin{equation}
j^{\mathrm{CC}}_\alpha = 2
\left(
\bar{e}_L \gamma_{\alpha} \nu_{eL}
+
\bar{\mu}_L \gamma_{\alpha} \nu_{\mu L}
+
\bar{\tau}_L \gamma_{\alpha} \nu_{\tau L}
\right)
\label{jCC}
\end{equation}
is the charged current;

\item
The Hamiltonian of neutral current (NC) weak interactions
\begin{equation}
\mathcal{H}_I^{\mathrm{NC}}
=
\frac{g}{2\cos{\vartheta_{\mathrm{W}}}} \ j^{\mathrm{NC}}_{\alpha} \ Z^\alpha
\,,
\label{HNC}
\end{equation}
where $\vartheta_{\mathrm{W}}$ is the weak (Weinberg) angle,
$Z^\alpha $ is the field of the neutral vector boson,
and

\begin{equation}
j^{\mathrm{NC}}_\alpha
=
\bar{\nu}_{eL} \gamma_\alpha \nu_{eL}
+
\bar{\nu}_{\mu L} \gamma_\alpha \nu_{\mu L}
+
\bar{\nu}_{\tau L} \gamma_\alpha \nu_{\tau L}
+
\ldots
\label{jNC}
\end{equation}
is the neutral current\footnote{Notice that
in the Standard Model the constants $g$,
$\sin{\vartheta_{\mathrm{W}}}$ and $e$
are not independent but rather are related by the unification condition
$ g \sin{\vartheta_{\mathrm{W}}} = e $ from which the masses of
the $W^{\pm}$ and $Z^0$ 
bosons can be inferred.
Indeed,
these predictions were confirmed by the data of high
precision LEP experiments
\cite{PDG98}.}.

\end{enumerate}

The three fields of
$\nu_e$, $\nu_\mu$ and $\nu_\tau$
enter in the standard charged and neutral currents (\ref{jCC}) and (\ref{jNC}).
There exist no light flavor neutrinos other than these.
This was impressively proved
by the measurement of the width of the decay
$ Z \to \nu + \bar{\nu} $
in the SLC and LEP experiments
\cite{PDG98}.
In the framework of the SM, the width is determined only by the number $N_\nu$ of
light flavor neutrinos.
In the recent LEP experiments, this number
was found to be
\begin{equation}
N_\nu = 2.994 \pm 0.011
\,.
\end{equation}

\begin{table}[t]
\begin{center}
\begin{tabular}{|c|c|c|c|} \hline
       & $ (\nu_e,~e^-) $ & $ (\nu_\mu,~\mu ^-) $ & $ (\nu_\tau,~\tau ^-) $ \\            
\hline
$ L_e    $ &   1     &     0       &      0         \\ 
$ L_\mu  $ &   0     &     1       &      0         \\ 
$ L_\tau $ &   0     &     0       &      1         \\  \hline
\end{tabular}
\end{center}
\caption{ \label{Lepton numbers}
Lepton numbers of neutrinos and charged leptons}
\end{table}

The three different types of neutrinos are distinguished by the 
values of three different conserved lepton numbers: 
electron $L_e$, muon $L_\mu$ and tau $L_\tau$
(see Table~\ref{Lepton numbers}).
Zero lepton numbers are assigned to quarks, $\gamma$, $W$, $Z$, etc..
The SM interactions (\ref{Hem}), (\ref{HCC}), (\ref{HNC})
conserve separately
the total electron, muon and tau lepton numbers:
\begin{equation}
\sum L_e = \mathrm{const}
\,,
\qquad
\sum L_\mu = \mathrm{const}
\,,
\qquad
\sum L_\tau = \mathrm{const}
\,.
\label{lepton}
\end{equation}
Up to now,
there is no indication of a violation of this conservation law 
in weak interaction processes
as weak decays, neutrino interactions, etc..
From the existing data, rather strong limits on the probabilities of
the processes that are forbidden by (\ref{lepton}) have been obtained:
\begin{eqnarray}
&&
\Gamma( \mu \to e \, \gamma ) / \Gamma( \mu \to \mathrm{all} )
<
4.9 \times 10^{-11}
\,,
\label{lim01}
\\
&&
\Gamma( \mu \to 3 \, e ) / \Gamma( \mu \to \mathrm{all} )
<
1.0 \times 10^{-12}
\,,
\label{lim02}
\\
&&
\sigma( \mu^- \, \mathrm{Ti} \to e^- \, \mathrm{Ti} )
 / \sigma( \mu^- \, \mathrm{Ti} \to \mathrm{capture} )
<
4.3 \times 10^{-12}
\,,
\label{lim03}
\\
&&
\Gamma( K_L \to e \, \mu ) / \Gamma( K_L \to \mathrm{all} )
<
3.3 \times 10^{-11}
\,,
\label{lim04}
\\
&&
\Gamma( K^+ \to \pi^+ \, e^- \, \mu^+ ) / \Gamma( K^+ \to \mathrm{all} )
<
2.1 \times 10^{-10}
\,,
\label{lim05}
\end{eqnarray} 
at 90\% CL
\cite{PDG98}.

In spite of the above impressive data,
modern gauge theories suggest that
\textit{the lepton number conservation law is only approximate}.
It is violated
if neutrinos are massive and mixed. 
In the next Section we will discuss neutrino masses 
and neutrino mixings with emphasis on how,
as a consequence,
the law of lepton number conservation is violated.

\section{Neutrino mass and mixings}
\label{Neutrino mass and mixings}

The fields of $d$, $s$, $b$ quarks with charge $-1/3$ (``down quarks'') 
enter in the charged current of the SM
in a mixed form. The quark charged current is 
\begin{equation}
j^{\mathrm{CC}}_{\alpha} = 2 \left(
\bar{u}_L \gamma_{\alpha} d'_L + 
\bar{c}_L \gamma_{\alpha} s'_L + 
\bar{t}_L \gamma_{\alpha} b'_L
\right)
\,,
\end{equation}
where
\begin{equation}
d'_L = \sum _{q=d,s,b} V_{uq} \ q_L
\,,
\qquad
s'_L = \sum _{q=d,s,b} V_{cq} \ q_L
\,,
\qquad
b'_L = \sum _{q=d,s,b} V_{tq} \ q_L
\,.
\end{equation}
Here $u$, $c$, $t$ are the fields of the quarks with charge $2/3$
(``up quarks'')
and $ V $ is the unitary Cabibbo-Kobayashi-Maskawa (CKM) mixing matrix
\cite{CKM}.

Quark mixing is a well-established phenomenon.
The values of the elements 
of the CKM matrix have been determined
from the results of many experiments and are known with a high accuracy
\cite{PDG98}.
Quark mixing is a possible source of the CP-violation observed in
the decays of neutral K-mesons
\cite{discovery-of-CP-violation}
through the phase which enters in the
CKM mixing matrix.

\textit{What about neutrinos}?
Are neutrinos massive 
and, if so,
do the fields of massive neutrinos,
like the fields of quarks,
enter into the lepton charged current in a mixed form?
The answer to these questions is of fundamental importance for particle physics.

The following upper bounds have been obtained
in the experiments on the direct measurement of the masses
of neutrinos:
\begin{eqnarray}
m_{\nu_e} \lesssim 3 \, \mathrm{eV}
&\qquad&
\mathrm{(90\% \, CL)}
\,,
\\
m_{\nu_\mu} < 170 \, \mathrm{keV}
&\qquad&
\mathrm{(90\% \, CL)}
\,,
\\
m_{\nu_\tau} < 18.2 \, \mathrm{MeV}
&\qquad&
\mathrm{(95\% \, CL)}
\,.
\end{eqnarray}
The upper bound for
the $\nu_e$ mass was obtained from the experiments on the
measurement of high energy part of the $\beta$-spectrum of
$^3$H-decay
\cite{Tritium}.
The upper bound for the $\nu_\mu$ mass comes from the 
experiments on the measurement of muon momentum in the
decay $\pi ^+ \to \mu ^+ \nu_\mu$
\cite{Assamagan96} and
the upper bound for the $\nu_\tau $
mass from experiments on the measurement of
the distribution of effective mass
of five pions in the decay $\tau \to \nu_\tau + 5\pi$
\cite{Barate98}.

Although these data do not exclude massless neutrinos,
there are no convincing reasons for massless neutrinos.
Moreover, in theories such as Grand Unified gauge theories, for example, 
it is very natural for neutrinos to be massive particles:
the non-conservation
of lepton numbers and the appearance of right-handed neutrino fields in the
Lagrangian are generic features of these theories.

From all the existing neutrino data and from the astrophysical constraints
\cite{PDG98}
we can expect, however, that neutrino masses
are small.
To probe hard-to-find effects of small neutrino masses,
it requires some very sensitive special experiments. 
Such experiments have turned out to be
\textit{neutrino oscillation experiments}.
Before we discuss neutrino
oscillations in detail in the following
Sections, we will first discuss 
different possibilities of mixing of massive neutrinos.
 
As mentioned already, the minimal Standard Model is based on the
\textit{assumption}
that neutrino fields are left-handed two-component fields and there are
no right-handed fields in the Lagrangian. 
In such a model, neutrinos are two-component massless particles
(as in Fig.~\ref{Helicities}).

Neutrino masses
can be generated, however, by the same standard Higgs
mechanism which generates the masses
of quarks and charged leptons,
due to Yukawa interactions of neutrinos with the Higgs boson.
This interaction requires not only the left-handed doublets
(\ref{left-handed doublets}),
but also
right-handed singlets
$\nu_{\ell R}$.
In this case, the neutrino mass term is given by
\begin{equation}
\mathcal{L}
=
-
\sum_{\ell',\ell}
\bar{\nu}_{\ell'R} \ M_{\ell'\ell} \ \nu_{\ell L} + \mathrm{h.c.}
\,,
\label{Dirac mass}
\end{equation}
where $M$ is a complex matrix.
If $M$ is non-diagonal,
this Lagrangian does not conserve the lepton numbers
and the flavor neutrino fields $\nu_{eL}$, $\nu_{\mu L}$, $\nu_{\tau L}$
are given by
\begin{equation}
\nu_{\ell L} = \sum_{i=1}^{3} U_{\ell i} \ \nu_{iL}
\qquad
( \ell = e, \mu, \tau )
\,.
\label{Dirac mixing}
\end{equation}
Here $\nu_i$ is the field of the neutrino with mass $m_i$
and $U$ is the unitary mixing matrix.

The mass term (\ref{Dirac mass}) does not conserve
the lepton numbers $L_e$, $L_\mu$ and $L_\tau$
separately, but it conserves the total lepton number
\begin{equation}
L = L_e + L_\mu + L_\tau
\label{total lepton number} 
\end{equation}
and the neutrinos $\nu_i$ with definite mass are
\textit{four component Dirac particles}
(neutrinos and antineutrinos have,
correspondingly, $L=1$ and $L=-1$).

If Dirac neutrino masses are generated by the same mechanism as quark and
charged lepton masses, the neutrino masses are
only additional parameters of the SM and there is no
rationale for 
the smallness of neutrino masses compared with the masses
of the corresponding charged leptons.

A mechanism that explains the smallness of neutrino masses
exists if the neutrinos $\nu_i$ with definite masses
are \textit{two-component Majorana particles}.
It is the famous see-saw mechanism of neutrino mass generation
\cite{see-saw}.
The see-saw mechanism is based on the assumption that all
lepton numbers 
are violated at a scale $M$ that is much larger than the electroweak scale 
(usually $M \sim 10^{16} \, \mathrm{GeV}$).
For the neutrino mixing, we have the same
expression as (\ref{Dirac mixing}),
but in this case the field of
neutrino with mass $m_i$ satisfies the \textit{Majorana condition}
\begin{equation}
\nu_i^c = \nu_i
\end{equation}
($\nu_i^c$ is the charge-conjugated field)
and the value of the neutrino mass $m_i$ is given by the see-saw relation
\begin{equation}
m_i \simeq \frac{(m_f^i)^2}{M}
\,.
\end{equation}
Here $m_f^i$ is the mass of up-quark or lepton in the $i^{\mathrm{th}}$-generation.

Massive Majorana neutrinos are truly neutral particles that have no
lepton charge (neutrinos are identical to antineutrinos).
Majorana neutrino masses can be generated 
only in the framework of models beyond the SM in which the conservation of the
total lepton number is violated.

If massive neutrinos are Majorana particles, the number of
massive light neutrinos can be larger than three (the number of flavor
neutrinos).
In this case,
we have 
\begin{equation}
\nu_{\ell L} = \sum_{i=1}^n U_{\ell i} \ \nu_{iL}
\,,
\qquad
\nu_{sL} = \sum_{i=1}^n U_{si} \ \nu_{iL}
\,,
\label{Majorana mixing}
\end{equation}
where $\nu_s$ is the charge conjugated right-handed field
($\nu_{sL} = {\nu_{sR}}^c$) and $n=3+n_s$,
where $n_s$ is the number of sterile neutrinos.

Right-handed neutrino
fields do not enter into the charged and neutral currents of the SM
(see Eqs.(\ref{HCC}) and (\ref{HNC})).
This means that the quanta
of the neutrino fields $\nu_s$
do not interact with matter
(via the standard weak interaction).
Such neutrinos are called
\textit{sterile neutrinos}.
Because of
neutrino mixing,
the flavor neutrinos $\nu_e$, $\nu_\mu$ and $\nu_\tau$ can
transform into sterile states. Such a possibility is widely discussed now in the
literature in order to accommodate all the existing
neutrino oscillation data.

Some important questions that
are currently under active investigation are:

\begin{enumerate}

\item
What are the values of neutrino masses?

\item
What are the values of 
the elements of the unitary neutrino mixing
matrix $U$ ? 

\item
What is the nature of massive neutrinos? Are they Dirac particles with
total lepton number 
equal to $1$ ($-1$)
or truly neutral Majorana particles
with zero lepton number?

\item
Are there transitions of active neutrinos into sterile states?

\item
Is CP violated in the lepton sector?

\end{enumerate} 

Many experiments designed to observe neutrino oscillations,
to investigate neutrinoless double $\beta$-decay
\begin{equation}
(A,Z) \to (A,Z + 2) + e^- + e^-
\,,
\end{equation}
and to perform a
precise measurement of the high-energy part of the $\beta$-spectrum of
$^3$H-decay are all aimed to answer these fundamental questions.
As we will see in the next Section, neutrino oscillations provide a
unique opportunity to reveal the
effects of extremely small neutrino masses and small mixing.

\section{Neutrino oscillations}
\label{Neutrino oscillations}

If neutrinos have small mass and are mixed particles, 
neutrino oscillations take place
\cite{no-revs-old,BGG98-review}.
Neutrino oscillations were considered 
as early as in 1957 by B. Pontecorvo
\cite{Pontecorvo-57-58}
and
flavor neutrino mixings has been discussed
in 1962 by Maki, Nakagawa and Sakata \cite{Maki62}.

In this Section, we will discuss neutrino oscillations in some detail.
From the quantum mechanical point of view,
neutrino oscillations are similar to the very well-known oscillations of neutral
kaons $ K^0 \rightleftarrows \bar{K}^0 $.
Let us consider a beam of neutrinos with momentum $p$.
In the case of neutrino mixing  
((\ref{Dirac mixing}) or (\ref{Majorana mixing})),
the state of a neutrino $|\nu_\ell\rangle$ ($ \ell = e, \mu ,\tau$)
produced in a weak process
(for example, $\pi \to \mu\nu_\mu$, $n\to pe^-{\bar{\nu}}_e$, etc.)
is a
\textit{coherent superposition} of the states of neutrinos with definite
mass,
\begin{equation}
|\nu_\ell\rangle = \sum_i U^*_{\ell i} \ |\nu_i\rangle
\,,
\end{equation}
where $|\nu_i\rangle$ is the state of
a neutrino with momentum $p$ and
energy
\begin{equation}
E_i = \sqrt{p^2 + m^2_i} \simeq {p + \frac{m_i^2}{2p}}
\,.
\end{equation}
At the time $t$ after production,
the neutrino is no longer described by a pure flavor state,
but by the state
\begin{equation}
|\nu_\ell\rangle_t = \sum _i U^*_{\ell i} \ e^{-iE_it} \ |\nu_i\rangle
\,.
\end{equation}
Neutrinos can only be detected via weak interaction processes. 
Decomposing the state $|\nu_\ell\rangle_t$ in terms of weak  eigenstates 
$|\nu_{\ell'}\rangle$, we have 
\begin{equation}
|\nu_\ell\rangle_t
=
\sum_{\ell'} |\nu_{\ell'}\rangle \ \mathcal{A}_{\nu_{\ell'};\nu_\ell}(t)
\,,
\end{equation}
where
\begin{equation}
\mathcal{A}_{\nu_{\ell'};\nu_\ell}(t)
=
\sum_i U_{\ell'i} \ e^{-iE_it} \ U_{\ell i}^*
\,.
\label{transition amplitude}
\end{equation}
Thus, if neutrino mixing takes place, 
the state of a neutrino produced at $t=0$
as a state with definite flavor
becomes at the time $t>0$ a superposition of
all possible states of flavor neutrinos.
The quantity $\mathcal{A}_{\nu_{\ell'};\nu_\ell}(t)$ is the
amplitude of the transition $\nu_\ell\to\nu_{\ell'}$ during the time $t$.
It is clear from Eq.(\ref{transition amplitude})
that the transitions among different neutrino flavors
are effects of the phase differences of the
different mass components of the flavor state.
For the transition probability we have\footnote{We use
natural units, $\hbar = c = 1$.}
\begin{equation}
P_{\nu_{\alpha}\to\nu_{\alpha'}} = 
\left|
\sum_i U_{\alpha'i}^* \ U_{\alpha i}
\left[ \exp\left(-i\ \frac{\Delta{m}^2_{i1} L}{2p}\right) - 1 \right]
+
\delta _{\alpha'\alpha}
\right|^2
\,,
\label{prob}
\end{equation}
where $\Delta{m}^2_{ij} \equiv m_i^2 - m_j^2 $ and $L \simeq t$
is the distance between the neutrino production and detection points.

The expression (\ref{prob}) is valid not only for
the transitions among flavor neutrinos
$\nu_e$, $\nu_{\mu}$,  $\nu_{\tau}$
but also for the transitions of
flavor neutrinos into sterile states $\nu_s$. In this case the
index $i$ runs over $n>3$ values ($n$ is the number of massive
neutrinos) and the indexes
$ \alpha' $,  $\alpha $ run over $ e $,  $\mu $,  $\tau $,  $s_1 $, $\ldots$.

As one can see from Eq.(\ref{prob}),
the transition probabilities depend on
the parameter $L/p$,
on $n-1$ neutrino mass squared differences
and on the elements of the neutrino mixing matrix,
that can be parameterized in terms of
$n(n-1)/2$ mixing angles
and
$(n-1)(n-2)/2$ phases.
If all $\Delta{m}^2_{i1}$ are so small that the inequalities 
\begin{equation}
\frac{\Delta{m}^2_{i1} L}{2p} \ll 1
\end{equation}
are satisfied,
we simply have $P_{\nu_{\alpha}\to\nu_{\alpha'}} = \delta_{\alpha\alpha'}$
and neutrino oscillations do not take place.

\begin{figure}[t]
\begin{center}
\includegraphics[bb=65 385 595 630,width=0.99\linewidth]{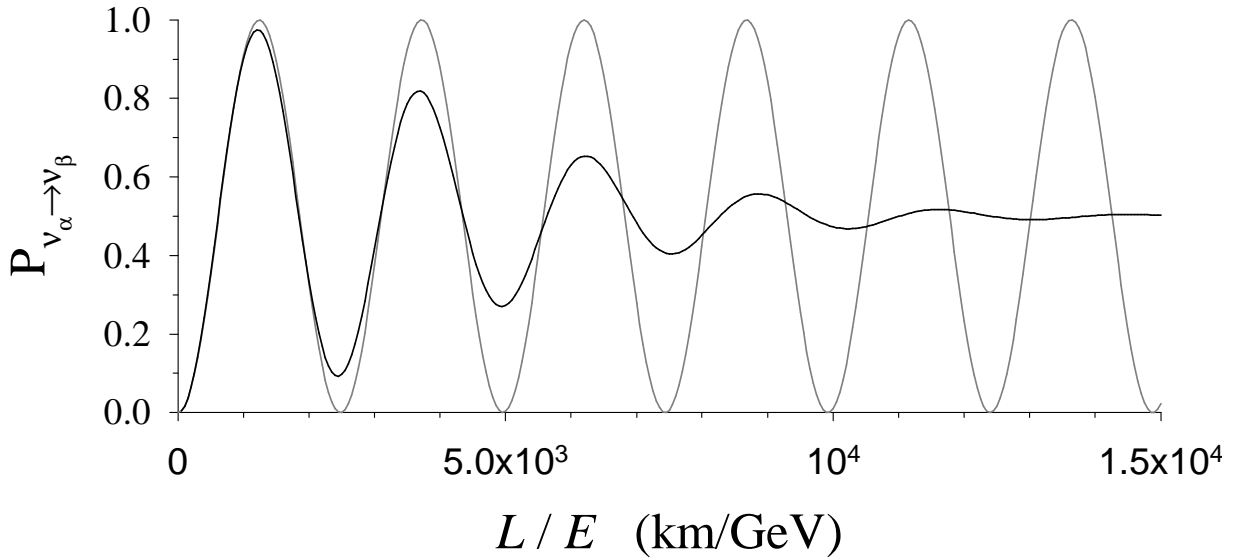}
\end{center}
\caption{ \label{proble}
Transition probability for $\sin^22\vartheta=1$
and $\Delta{m}^2=10^{-3} \, \mathrm{eV}^2$
as a function of
$ L / E $
in km/GeV (or m/MeV).
The grey line represents the transition probability (\ref{trans2})
and the black line represents the same transition probability
averaged over a Gaussian energy spectrum with
a mean energy value $\overline{E}=E$ and
standard deviation $\sigma = E/10$.}
\end{figure}

Let us consider the simplest case of mixing of two types of neutrinos.
In this case we have
\begin{eqnarray}
&&
\nu_{\alpha L}
=
\cos{\vartheta} \ \nu_{1L} + \sin{\vartheta} \ \nu_{2L}
\nonumber
\\
&&
\nu_{\alpha'L}
=
- \sin{\vartheta} \ \nu_{1L} + \cos{\vartheta} \ \nu_{2L}
\,,
\end{eqnarray}
where $\vartheta$ is the  mixing angle and $\alpha$, $\alpha'$ take the
values
$\mu, e$, or
$\mu, \tau$, or $e, \tau$, or $\mu, s$, etc..
From Eq.(\ref{prob}),
the transition probabilities are given by
\begin{eqnarray}
&&
P_{\nu_{\alpha}\to\nu_{\alpha'}} = 
\frac{1}{2} \ \sin^2{2\vartheta}
\left( 1 - \cos \frac{\Delta{m}^2 L}{2p} \right)
\qquad \qquad
(\alpha' \neq \alpha)
\label{trans1}
\\
&&
P_{\nu_{\alpha}\to\nu_{\alpha}}=
1 - P_{\nu_{\alpha}\to\nu_{\alpha'}} = 
1 -
\frac{1}{2} \ \sin^2{2\vartheta}
\left( 1 - \cos \frac{\Delta{m}^2 L}{2p} \right)
\label{surv1}
\,,
\end{eqnarray}
where $\Delta{m}^2 \equiv m^2_2 - m^2_1$.
The probability
$P_{\nu_{\alpha}\to\nu_{\alpha'}}$ can also be written in the form
\begin{equation}
P_{\nu_{\alpha}\to\nu_{\alpha'}}
=
\frac{1}{2} \ \sin^2{2\vartheta}
\left[ 1 - \cos\left( 2.48 \ \frac{\Delta{m}^2 L}{E} \right) \right]
\,,
\label{trans2}
\end{equation}
where $\Delta{m}^2$ is the neutrino mass squared difference in units of
eV$^2$,
$L$ is the distance in m (km) and $E$ is neutrino energy in MeV (GeV).
The mixing parameter $\sin^2{2\vartheta}$ is the oscillation amplitude.
Thus, for a fixed energy, the transition probability is a periodical function 
of the distance. The oscillation length $L_{\mathrm{osc}}$ that characterizes
this periodicity is given by 
\begin{equation}
L_{\mathrm{osc}} = 4\pi \ \frac{E}{\Delta{m}^2}
=
2.54 \ \frac{E \ (\mathrm{MeV})}{\Delta{m}^2 \ (\mathrm{eV}^2)} \ \mathrm{m}
\,.
\end{equation}
The transition probabilities among different neutrino
flavors depend on the parameter $L/E$.
This oscillatory behavior is shown in
Fig.~\ref{proble} for $\sin^22\vartheta=1$ and $\Delta{m}^2=10^{-3} \, \mathrm{eV}^2$
(grey line).

In practice, neutrino beams are not monoenergetic  
and neutrino sources and detectors have always a finite size.
The black curve in Fig.~\ref{proble} shows the effects of averaging of the transition
probability over a Gaussian neutrino spectrum
with a mean energy value $\overline{E}=E$
and
a standard deviation $\sigma = E/10$.

In order to observe neutrino oscillations 
it is necessary that
the neutrino mass squared difference 
$\Delta{m}^2$
satisfies the condition
\begin{equation}
\Delta{m}^2 \gtrsim \frac{E}{L}
\,.
\end{equation}
This condition provides a guideline for finding the sensitivity of 
neutrino oscillation
experiments to the neutrino mass squared difference
(for large values of
$\sin^2{2\vartheta}$). 
For example,
for reactor ($L \sim 100 \, \mathrm{m}$, $E \sim 1 \, \mathrm{MeV}$),
accelerator ($L \sim 1 \, \mathrm{km}$, $E \sim 1 \, \mathrm{GeV}$)
and
solar ($L \sim 10^{11} \, \mathrm{m}$, $E \sim 1 \, \mathrm{MeV}$) 
neutrino oscillation experiments,
the minimal values of $\Delta{m}^2$ are
$\sim 10^{-2} \, \mathrm{eV}^2$,
$\sim 1 \, \mathrm{eV}^2$, 
$\sim 10^{-11} \, \mathrm{eV}^2$,
respectively. 
However,
in order to calculate precisely the sensitivity
of neutrino oscillation experiments, it is necessary
to take into account also all the conditions of an experiment.
    
\begin{figure}[t]
\begin{center}
\includegraphics[bb=107 199 566 665,height=7cm]{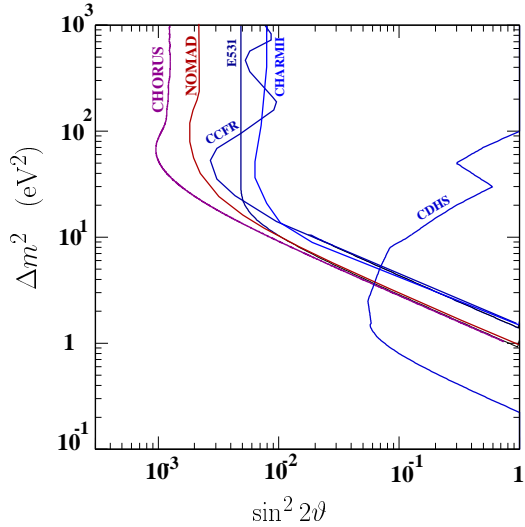}
\end{center}
\caption{ \label{nomad}
Exclusion curves (90\% CL) in the
$\nu_\mu\to\nu_\tau$
channel
obtained in the
CDHS \protect\cite{CDHS84},
FNAL E531 \protect\cite{FNALE531},
CHARM II \protect\cite{CHARMII94},
CCFR \protect\cite{CCFR95},
CHORUS \protect\cite{CHORUS}
and
NOMAD \protect\cite{NOMAD,NOMADb}
experiments.}
\end{figure}

A typical exclusion plot
in the $\sin^22\vartheta$--$\Delta{m}^2$ plane,
obtained from the data of experiments in which neutrino oscillations
were not found,
is presented in Fig.~\ref{nomad} \cite{NOMADb}.
This plot shows the exclusion curves in the
$\nu_\mu\to\nu_\tau$
channel
obtained in the
CDHS \cite{CDHS84},
FNAL E531 \cite{FNALE531},
CHARM II \cite{CHARMII94},
CCFR \cite{CCFR95},
CHORUS \cite{CHORUS}
and
NOMAD \cite{NOMAD,NOMADb}
accelerator experiments.
The excluded region lies on the right of the curves.

The expressions (\ref{trans1}) and (\ref{surv1})
that describe neutrino oscillations
between two types of neutrinos 
are usually employed in the analysis of experimental data. 
The general expressions for the transition probabilities among
three types of
neutrinos are rather complicated, but the
probabilities become
simple if there is a hierarchy of neutrino masses,
\begin{equation}
m_1 \ll m_2 \ll m_3
\,.
\end{equation}
Such a hierarchy is realized, for example, if neutrino masses are
generated by the see-saw mechanism.

In the case of a mass hierarchy
the
$\nu_\alpha \to \nu_\alpha'$ transition probability
and the
$\nu_\alpha \to \nu_\alpha$ survival probability
in neutrino oscillation experiments for which only the largest mass-squared
$\Delta{m}^2_{31}$
is relevant
are given by
\cite{one-mass-old}
\begin{eqnarray}
&&
P_{\nu_{\alpha}\to\nu_{\alpha'}}
=
\frac{1}{2} \ A_{\alpha;\alpha'}
\left( 1 - \cos \frac{\Delta{m}^2_{31} L}{2p} \right)
\qquad \qquad
( \alpha' \neq \alpha )
\,,
\label{trans3}
\\
&&
P_{\nu_{\alpha}\to\nu_{\alpha}}
=
1 - \frac{1}{2} \ B_{\alpha;\alpha}
\left( 1 - \cos \frac{\Delta{m}^2_{31} L}{2p} \right)
\,,
\label{surv3}
\end{eqnarray}
with the oscillation amplitudes
\begin{equation}
A_{\alpha;\alpha'} = 4 \ |U_{\alpha3}|^2 \ | U_{\alpha'3}|^2
\,,
\qquad
B_{\alpha;\alpha} = 4 \ |U_{\alpha3}|^2 \ (1-|U_{\alpha3}|^2)
\,.
\end{equation}
The expressions (\ref{trans3}) and (\ref{surv3})
have the same form as the two-neutrino
expressions (\ref{trans1}) and (\ref{surv1}),
respectively.
They describe, however, all possible
transitions among three types of neutrinos:
$\nu_\mu\leftrightarrows\nu_e$,
$\nu_\mu\leftrightarrows\nu_\tau$
and
$\nu_e\leftrightarrows\nu_\tau$.

The transition probabilities (\ref{trans3}) and (\ref{surv3})
are characterized by the two mixing
parameters $|U_{e3}|^2$ and $|U_{\mu3}|^2$
(from the unitarity of the mixing matrix it follows that
$|U_{\tau3}|^2 = 1 - |U_{e3}|^2 - |U_{\mu3}|^2$)
and by one neutrino mass-squared
difference $ \Delta{m}^2_{31} \equiv m_3^2 - m_1^2 $.
The expressions (\ref{trans3}) and (\ref{surv3})
are currently often used in the analysis of 
data of reactor, accelerator and atmospheric neutrino oscillation experiments
\cite{one-mass-new}.

There are two types of oscillation experiments:

\begin{enumerate}

\item
Appearance experiments;

\item
Disappearance experiments.

\end{enumerate}

In the experiments of the first type, 
neutrinos
of a certain flavor (for example, $\nu_\mu$)
are produced and then the appearance of neutrinos of a
\textit{different flavor}
(for example, $\nu_\tau$)
are searched for at some distance.
In the experiments of the second type, 
neutrinos of a certain flavor
(say, $\nu_\mu$) are produced and, at some distance, 
neutrinos of the \textit{same flavor}
($\nu_\mu$) are detected.
In the latter case, if the
number of
detected neutrino events is less
than the expected number
(with the assumption that there are no oscillations),
one has a signal
that some neutrinos are transformed into neutrinos of other flavors.
Reactor neutrino experiments
in which
$\bar\nu_e\to\bar\nu_e$ transitions are searched for are typical 
disappearance experiments.
Accelerator neutrino oscillation experiments can be both of appearance and
of disappearance type.

In the next Section we will discuss some results of neutrino oscillation experiments.

\section{Status of neutrino oscillations}
\label{Status of neutrino oscillations}

The first experimental information on neutrino oscillations
have been obtained about twenty years ago
as a by-product of an experiment designed for other measurements.
At present, the majority of neutrino experiments are
dedicated to the detection of neutrino oscillations.

An impressive evidence
in favor of oscillations of atmospheric neutrinos has been obtained recently
in the Super-Kamiokande experiment
\cite{SK-atm-98,Scholberg-Venice-99,%
Kajita-Ringberg-99,Learned-JHU-99,Nakahata-TAUP-99}. 
The results of this experiment reported at the Neutrino '98 conference
\cite{nu98}
attracted enormous attention to the problem of neutrino mass
from the general public as well as from many
physicists\footnote[2]
{After the publication of the Super-Kamiokande data
more than 500 papers on the neutrino mass problem appeared
in the hep-ph electronic archive at
xxx.lanl.gov.}.

Indications in favor of neutrino oscillations have been obtained also in the 
Kamiokande \cite{Kam-atm},
IMB \cite{IMB},
Soudan 2 \cite{Soudan2}
and
MACRO \cite{MACRO}
atmospheric neutrino experiments.
In all the solar neutrino experiments
(Homestake \cite{Homestake},
Kamiokande \cite{Kam-sun},
GALLEX \cite{GALLEX},
SAGE \cite{SAGE},
Super-Kamiokande \cite{SK-sun,Nakahata-TAUP-99})
the observed 
event rates are significantly smaller than the expected ones.
This \textit{solar neutrino problem} can naturally be explained
if neutrinos are massive and mixed.
Indications in favor of neutrino oscillations were also found in the accelerator
LSND neutrino experiment \cite{LSND-96-PRL,LSND-98-PRL}.
In the rest of the accelerator neutrino experiments, no
indications in favor of neutrino oscillations have been found.
The reactor neutrino experiments have all failed to observe neutrino oscillations.

We will start with the discussion of the results of \textit{atmospheric neutrino
experiments}. The main source of atmospheric neutrinos is
the chain of decays
\begin{eqnarray}
\pi \to & \mu & \null + \nu_\mu
\nonumber
\\
& \downarrow &
\nonumber
\\
& e & \null + \nu_e + \nu_\mu
\,,
\label{chain}
\end{eqnarray}
pions being produced in the interaction of cosmic rays with the
atmosphere (in the currently running experiments, 
neutrinos and
antineutrinos are not distinguishable).
Almost all the muons with relatively low energies
($ \lesssim 1 \, \mathrm{GeV} $) 
have enough time to decay in the atmosphere, so that the ratio
of the fluxes of low energy muon and electron neutrinos is approximately equal
to two. At higher energies this ratio becomes larger.
 
The absolute fluxes of muon and electron neutrinos can be calculated with an
accuracy of 20--30\%.
However,
because of an approximate cancellation of the uncertainties of the absolute fluxes,
the ratio of the fluxes of muon and
electron neutrinos is predicted with an uncertainty of about 5\%
\cite{Gaisser-5pc}.
The results of atmospheric
neutrino experiments are usually presented in terms of the double ratio
of
the ratio of observed muon and electron events
and
the ratio of muon and electron events calculated with a Monte Carlo
under the assumption that there are no neutrino oscillations:
\begin{equation}
R
=
\left(\frac{N_\mu}{N_e}\right)_{\mathrm{obs}}
\
\Bigl/
\
\left(\frac{N_\mu}{N_e}\right)_{\mathrm{MC}}
\,.
\end{equation}
In four experiments
(Kamiokande \cite{Kam-atm},
IMB \cite{IMB},
Soudan 2 \cite{Soudan2}
and
Super-Kamiokande
\cite{SK-atm-98,Scholberg-Venice-99,%
Kajita-Ringberg-99,Learned-JHU-99,Nakahata-TAUP-99})
the observed values of the double ratio
$R$  significantly less than one.

In the Kamiokande, IMB and Super-Kamiokande experiments water Cherenkov
detectors are used.
The Super-Kamiokande detector is a huge tank filled with 50 kton of water
and covered with 11000 photo-multiplier tubes.
The Soudan2 detector is an iron calorimeter.

The Kamiokande and
Super-Kamiokande collaborations divide their events into two categories:
sub-GeV events with
$E_{\mathrm{vis}} < 1.33 \, \mathrm{GeV}$
and multi-GeV events with
$E_{\mathrm{vis}} > 1.33 \, \mathrm{GeV}$
($E_{\mathrm{vis}}$ is the visible energy). 
In the high statistics Super-Kamiokande experiment,
the double ratio $R$ was found to be \cite{Nakahata-TAUP-99}
\begin{eqnarray}
R = 0.680 \, {}^{+0.023}_{-0.022} \pm 0.053
& \qquad &
\mbox{(sub-GeV)}
\,,
\nonumber\\
R = 0.678 \, {}^{+0.042}_{-0.039} \pm 0.080
& \qquad &
\mbox{(multi-GeV)}
\,.
\end{eqnarray}
In other experiments, the double ratio R was found to be
\begin{eqnarray}
R = 0.60 \, {}^{+0.07}_{-0.06} \pm 0.05
& \qquad &
\mbox{(Kamiokande sub-GeV) \protect\cite{Hir92}}
\,,
\nonumber\\
R = 0.57 \, {}^{+0.08}_{-0.07} \pm 0.07
& \qquad &
\mbox{(Kamiokande multi-GeV) \protect\cite{Kam-atm}}
\,,
\nonumber\\
R = 0.54 \pm 0.05 \pm 0.11
& \qquad &
\mbox{(IMB) \protect\cite{IMB}}
\,,
\nonumber\\
R = 0.68 \pm 0.11 \pm 0.06
& \qquad &
\mbox{(Soudan 2) \protect\cite{Soudan2}}
\,.
\end{eqnarray}
The fact that the double ratio $R$ is less than one could mean the
disappearance of $\nu_\mu$ or appearance of $\nu_e$ or both. The
Super-Kamiokande collaboration found a compelling evidence in favor of
disappearance of $\nu_\mu$ in the multi-GeV region. 

A relatively large statistics of events allowed the Super-Kamiokande
Collaboration to
investigate in detail the zenith angle ($\theta$) dependence of the
number of electron and muon events. Down-going neutrinos
($\cos\theta\simeq1$) pass through a distance of about 20 km. Up-going
neutrinos 
($\cos\theta\simeq-1$) travel a distance of about 13000 km.
The Super-Kamiokande collaboration  observed a significant up-down
asymmetry of muon events in the multi-GeV region \cite{Learned-JHU-99}:
\begin{equation}
A_\mu ={\left(\frac{U-D}{U+D}\right)}_\mu
=
- 0.311 \pm 0.043 \pm 0.01
\,,
\end{equation}
where $U$ is the number of up-going events
with $\cos\theta$ in the range $-1 < \cos\theta < -0.2$ and $D$ is the
number of down-going events
with $\cos\theta$ in the range
$0.2 < \cos\theta < 1$. 
Thus, the up-down asymmetry of multi-GeV muon events deviates from zero
by about seven standard deviations. On the other hand,
the asymmetry of
electron events
is consistent with zero \cite{SK-atm-98}:
\begin{equation}
A_e = -0.036\pm 0.067\pm 0.02
\,.
\end{equation}
The negative sign
of the asymmetry $A_\mu$ means that the number of up-going muon events is
smaller than
that of down-going events. The $U/D$ ratio is given by
\cite{Kajita-Ringberg-99}
\begin{eqnarray}
{\left(\frac{U}{D}\right)}_\mu = 0.52 \, {}^{+0.05}_{-0.04} \pm 0.01
\,.
\end{eqnarray}
The disappearance of up-going muon neutrinos 
can be naturally explained by 
$\nu_\mu \to \nu_\tau$ oscillations,
since the up-going $\nu_\mu$'s travel a much longer distance than the down-going
$\nu_\mu$'s. 

In the case of oscillations between two types of neutrinos, the transition
probability depends on two parameters: $\Delta{m}^2$ and
$\sin^22\vartheta$.
From the analysis of the Super-Kamiokande data, the best-fit
values of these parameters were found to be \cite{Nakahata-TAUP-99}
\begin{equation}
\Delta{m}^2 \simeq 3 \times 10^{-3} \, \mathrm{eV}^2
\,,
\qquad
\sin^2{2\vartheta} \simeq 1
\,.
\end{equation}

In the case of quarks, all the mixing angles are known to be small.
The Super-Kamiokande result shows
that neutrino mixings are very different from quark mixings: the 
mixing angle that characterizes $\nu_\mu \to \nu_\tau$ transitions
inferred from the atmospheric neutrino data is large (close to $\pi/4$).

\begin{figure}[t]
\begin{center}
\includegraphics[bb=5 0 505 485,height=7cm]{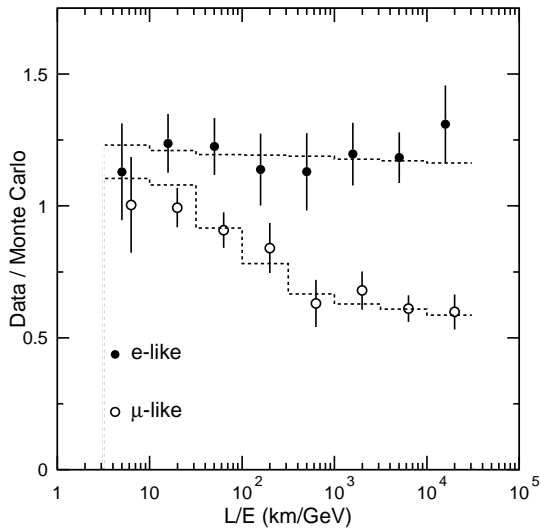}
\end{center}
\caption{ \label{loe}
The ratio of the number of data events to Monte Carlo calculated events
(in the absence of oscillations)
in Super-Kamiokande,
as a function of reconstructed $L/E$
\protect\cite{SK-atm-98}.
The points show the ratio of observed data and Monte Carlo expectation.
The dashed lines show the expected ratios
for $\nu_\mu\to\nu_\tau$
(or $\nu_\mu\to\nu_s$)
oscillations
with
$ \Delta{m}^2 = 2.2 \times 10^{-3} \, \mathrm{eV}^2 $
and
$ \sin^22\vartheta = 1 $.
The slight $L/E$ dependence
of the ratio for $e$-like events
is due to contamination of $\nu_\mu$ CC interactions (2--7\%).}
\end{figure}

The neutrino transition probabilities also depend on the parameter $L/E$.
In Fig.~\ref{loe}
the ratio of the numbers of observed and predicted muon (electron) events
as a function of $L/E$ is shown.
The ratio practically does not depend on $L/E$ for the electron events,
but strongly depends on $L/E$ for the muon events. In the region
$L/E \gtrsim 10^3 \mathrm{km}/\mathrm{GeV}$
the argument of the cosine in the expression for the 
$\nu_\mu$ survival probability (see Eq.(\ref{surv1}))
is large and the cosine in the $\nu_\mu$ survival probability
disappears due to averaging over energies and distances. As
a result, in this region  we have
$
\langle P_{\nu_\mu\to\nu_\mu} \rangle
\simeq
1-\frac{1}{2}\sin^2{2\vartheta}
\simeq
\frac{1}{2}
$
(see the last four points in Fig.~\ref{loe}).
The disappearance of atmospheric
muon neutrinos can also be explained by $\nu_\mu \to \nu_s$
oscillations. These two alternatives can be distinguished
with the observation of atmospheric neutrinos through the NC process
\cite{K2K,Vis98a}
\begin{equation}
\nu_\ell + A \to \nu_\ell + \pi^0 + X
\qquad
(\ell=e,\mu,\tau)
\,.
\end{equation}
If $\nu_\mu$'s transform into sterile states, an up-down asymmetry
of $\pi^0$ events should be observed.
Another possibility to distinguish the
$\nu_\mu\to\nu_\tau$
and
$\nu_\mu\to\nu_s$
channels is to look for matter effects.
The investigation of both effects
allowed the Super-Kamiokande Collaboration to exclude
pure $\nu_\mu\to\nu_s$ transitions
at $2\sigma$ level
\cite{Kajita-Ringberg-99,Learned-JHU-99,Nakahata-TAUP-99}.

The value $\Delta{m}^2 \simeq 10^{-3} \, \mathrm{eV}^2$,
which is just right in explaining  
the atmospheric neutrino data,
can be probed in the long-baseline (LBL) reactor and accelerator
experiments with a distance between source and detector about 1 km in
the case of reactors and about 1000 km in the case of accelerators.
No indications in favor of 
neutrino oscillations were found in
the first reactor LBL experiment
CHOOZ \cite{CHOOZ}.
The CHOOZ data exclude
transitions of
$\bar\nu_e$ into all other possible antineutrinos for
$\Delta{m}^2 \gtrsim 10^{-3} \, \mathrm{eV}^2$ and  
$\sin^2{2\vartheta} \gtrsim 0.1$.

The first accelerator LBL experiment K2K,
from KEK to Super-Kamiokande
(a distance of about 250 km),
started in Japan in 1999 \cite{K2K}.
The Fermilab--Soudan (a distance of about 730 km) 
LBL experiment
MINOS \cite{MINOS} and
the CERN-Gran Sasso
(a distance of about 730 km)
program of LBL experiments
\cite{CERN-LNGS}
will start after the year 2000.
These experiments will investigate in detail the transitions of
accelerator $\nu_\mu$ 's into all possible states in the atmospheric neutrino
region of $\Delta{m}^2$.

We will now discuss the results of \textit{solar neutrino experiments}.
The energy of the Sun is produced in the reactions of the thermonuclear $pp$ chain 
and CNO cycle.
From the thermodynamical point of view, the source of the energy of the Sun
is a transformation of four protons into $^4$He: 
\begin{equation}
4p \to {}^4\mathrm{He} + 2e^+ + 2\nu_e
\,.
\label{4p}
\end{equation}
Thus, the production of the energy of the Sun \textit{is accompanied} by
the emission of electron neutrinos. 

\begin{table}[t]
\begin{center}
\renewcommand{\arraystretch}{1.45}
\begin{tabular}{|c|c|c|}
\hline
reaction
&
\begin{tabular}{c}
neutrino energy
\\
(MeV)
\end{tabular}
&
\begin{tabular}{c}
expected flux \protect\cite{BP98}
\\
(cm$^{-2}$ s$^{-1}$)
\end{tabular}
\\
\hline
$ p + p \to d + e^+ + \nu_e $
&
$ \leqslant 0.42 $
&
$ 5.9 \times 10^{10} $
\\
$ e^- + {}^{7}\mathrm{Be} \to {}^{7}\mathrm{Li} + \nu_e $
&
$ 0.86 $
&
$  4.8 \times 10^{9}$  \\
$ {}^8\mathrm{B} \to {}^8\mathrm{Be}^* + e^+ + \nu_e $
&
$ \lesssim 15 $
&
$ 5.2 \times 10^{6}$
\\
\hline
\end{tabular}
\end{center}
\caption{ \label{sources}
Main sources of solar $\nu_e$'s.}
\end{table}

The main sources of $\nu_e$ are the reactions of the $pp$ chain that are listed
in Table~\ref{sources},
and the neutrinos coming from these sources are called
$pp$, $^7$Be and $^8$B neutrinos, respectively.
Neutrinos from other sources are called
$pep$,
$hep$
(from the reactions
$ p + e^- + p \to d + \nu_e $,
$ {}^3\mathrm{He} + p \to {}^4\mathrm{He} + e^+ + \nu_e $
of the $pp$ chain)
and
$^{13}$N,
$^{15}$O,
$^{17}$F
(from the reactions
$ {}^{13}\mathrm{N} \to {}^{13}\mathrm{C} + e^+ + \nu_e $,
$ {}^{15}\mathrm{O} \to {}^{15}\mathrm{N} + e^+ + \nu_e $,
$ {}^{17}\mathrm{F} \to {}^{17}\mathrm{O} + e^+ + \nu_e $
of the CNO cycle)
neutrinos.
As one can see from Table~\ref{sources}, the major part of solar neutrinos are low
energy $pp$ neutrinos with $E \leqslant 0.42 \, \mathrm{MeV}$.
There are about 10\% of
monoenergetic $^7$Be neutrinos with an energy of 0.86 MeV,
whereas
the high energy $^8$B neutrinos
($E \lesssim 15 \, \mathrm{MeV}$)
constitute a very small part of the total flux of
$\nu_e$'s from the Sun.
However, as we will see later, these neutrinos
give the major contribution to the event rates of experiments
with a high threshold.  

Assuming that $P_{\nu_e\to\nu_e}^{\mathrm{sun}} = 1$
and that the Sun is in a stable state,
the transition in Eq.(\ref{4p})
implies the following general relation between
the neutrino fluxes and the luminosity of the Sun
\cite{PDG98}
$ \mathcal{L}_\odot = 2.40 \times 10^{39} \, \mathrm{MeV} \, \mathrm{s}^{-1} $:
\begin{equation}
\sum_r
\left( \frac{Q}{2} - \overline{E}_r \right) \Phi_r
=
\frac{{\mathcal{L}}_\odot}{4\pi R^2}
\,,
\label{luminosity}
\end{equation}
with
$
r = pp , \, {}^7\mathrm{Be} , \, {}^8\mathrm{B} , \, pep , \, hep , \,
{}^{13}\mathrm{N} , \, {}^{15}\mathrm{O} , \, {}^{17}\mathrm{F}
$.
Here $Q = 26.7 \, \mathrm{MeV}$ is the energy release in
the transition (\ref{4p}),
$ R = 1.50 \times 10^{13} \, \mathrm{cm} $
is the Sun--Earth distance,
and $\Phi_r$ and $\overline{E}_r$ are the total flux
and the average energy of neutrinos
from the source $r$, respectively. 

Now let us turn to the experimental data.
The pioneering Chlorine solar neutrino
experiment by R. Davis \textit{et al.}
\cite{Homestake} known as the Homestake experiment
started more than 30 years ago.
Now the results of five solar neutrino experiments  
are available. These results are presented in Table~\ref{solar}.

Homestake \cite{Homestake},
GALLEX \cite{GALLEX}
and
SAGE \cite{SAGE}
are underground radiochemical experiments.
The target in the Homestake experiment is a tank 
filled with 615 tons of C$_2$Cl$_4$ liquid.
Solar neutrinos are detected in this experiment through the extraction of
radioactive $^{37}$Ar atoms  
produced in the Pontecorvo-Davis reaction
\begin{equation}
\nu_e + {}^{37}\mathrm{Cl} \rightarrow e^- + {}^{37}\mathrm{Ar}
\,.
\label{Cl-Ar}
\end{equation}
The atoms of $^{37}$Ar are extracted
from the tank by purging it with $^4$He gas and 
the Auger electrons
produced in the capture of $e^-$ by 
$^{37}$Ar are detected in a low-background proportional counter.
During a typical exposure time of 2 months
about 16 atoms of $^{37}$Ar
are extracted from the volume that contains $2.2\times10^{30}$ atoms of $^{37}$Cl!

The observed event rate in the Homestake experiment
averaged over 108 runs is $2.56 \pm 0.23$ SNU,
whereas the event rate predicted 
by the Standard Solar Model (SSM) \cite{BP98} is
$7.7 \pm 1.2 $ SNU.

The threshold of
the reaction (\ref{Cl-Ar})
is $E_{\mathrm{th}} = 0.81 \, \mathrm{MeV}$. 
Thus, low energy $pp$ neutrinos cannot be detected in the 
Homestake
experiment. 
The main contributions to the event rate
come from $^8$B and $^7$Be neutrinos (according to the SSM,
77\% and 15\%, respectively).

\begin{table}[t]
\begin{center}
\renewcommand{\arraystretch}{1.45}
\begin{tabular}{|c|c|c|} \hline
Experiment & Observed rate        & Predicted rate \protect\cite{BP98}
\\
\hline
Homestake
&
$ 2.56 \pm 0.16 \pm 0.16 $
&
$ 7.7 \, {}^{+1.2}_{-1.0} $
\\     
GALLEX
&
$ 77.5 \pm 6.2 \, {}^{+4.3}_{-4.7} $
&
$ 129 \, {}^{+8}_{-6} $
\\ 
SAGE
&
$ 66.6 \, {}^{+6.8}_{-7.1} \, {}^{+3.8}_{-4.0} $
&
"
\\
Kamiokande
&
$ ( 2.80 \pm 0.19 \pm 0.33 ) \times 10^{6} $
&
$ ( 5.15 \, {}^{+1.0}_{-0.7} ) \times 10^{6} $
\\
Super-Kamiokande
&
$ ( 2.45 \pm 0.04 \pm 0.07 ) \times 10^{6} $
&
"
\\   
\hline
\end{tabular}
\end{center}
\caption{ \label{solar}
The results of solar neutrino experiments.
The rates measured in Homestake, GALLEX and SAGE are given in SNU
($1\,\mathrm{SNU}=10^{-36} \, \mathrm{events/atom/sec}$).
The rates measured in Kamiokande and Super-Kamiokande are given in
$ \mathrm{cm}^{-2} \mathrm{s}^{-1} $.}
\end{table}

Neutrinos from all the reactions
in the Sun are detected
in the radiochemical Gallium experiments GALLEX 
(30.3 tons of  $^{71}$Ga in gallium-chloride solution)
and SAGE (57 tons of $^{71}$Ga in metallic form).
The detection is based on 
the observation of radioactive $^{71}$Ge atoms produced in the
reaction
\begin{equation}
\nu_e + {}^{71}\mathrm{Ga} \rightarrow e^- + {}^{71}\mathrm{Ge}
\,,
\end{equation}
whose threshold is $E_{\mathrm{th}}=0.23\,\mathrm{MeV}$. The event rates
observed in the
GALLEX and SAGE experiments are about 1/2 of the predicted rates
(see Table~\ref{solar}).

In the underground
Kamiokande \cite{Kam-sun}
and
Super-Kamiokande \cite{SK-sun,Nakahata-TAUP-99}
water-Cherenkov experiments,
the solar neutrinos are detected in real time through the observation of
the recoil electron in the process
\begin{equation}
\nu + e \to \nu + e
\,.
\end{equation}
Most importantly, in these experiments \textit{the direction of neutrinos can
be determined} through the
the measurement of the direction of the recoil electrons. 
Because of the low energy background 
from natural radioactivity,  the
thresholds in the
Kamiokande and Super-Kamiokande experiments are rather high:
$E_{\mathrm{th}} \simeq 7 \, \mathrm{MeV}$ and 
$E_{\mathrm{th}} \simeq 5.5 \, \mathrm{MeV}$, respectively.
Thus, in these experiments only $^8$B neutrinos are detected.

The results of the Kamiokande and Super-Kamiokande experiments are presented
in Table~\ref{solar}.
As it can be seen from this Table, the 
$^8$B neutrino flux observed in these experiments is significantly smaller (about 1/2)
than the SSM prediction.

Thus, in all the solar neutrino experiments a \textit{deficit of solar
$\nu_e$'s
is
observed}. This deficit constitutes the solar neutrino problem.
What is the origin of the problem?

The predictions of the SSM are considered as rather robust
\cite{BP98,Bahcall-nu98}.
The model 
takes into account all the existing data on nuclear cross sections,
opacities, etc. and is
in impressive agreement with precise helioseismological data. 
It is a consensus that the solar neutrino
problem should be attributed to
\textit{neutrino properties}.

Let us first discuss some model-independent conclusions that can be inferred 
from the solar neutrino data.
In the gallium experiments, neutrinos from all  the solar sources are detected.
From the luminosity constraint (\ref{luminosity}), we have 
the following lower bound for the
event rate in the gallium experiments:
\begin{equation} 
S_{\mathrm{Ga}}
=
\int \mathrm{d}E \ \sigma_{\mathrm{Ga}}(E)
\sum_r \Phi_r(E)
\geqslant
76 \pm 2 \ \mathrm{SNU}
\,.
\end{equation}
The data of the GALLEX and SAGE experiments are compatible with this bound
(see the Table~\ref{solar}). 
Thus, on the basis of the luminosity constraint alone, 
one cannot
conclude that the solar neutrino problem exists. However,
we can see that the problem exists if we \textit{compare} the data obtained from
different solar neutrino
experiments
\cite{Bahcall-Krastev-Smirnov-98,BGG98-review}.

Let us assume that
$P_{\nu_e\to\nu_e}^{\mathrm{sun}}=1$ and let us consider the total neutrino fluxes
$\Phi_r$ as free variable parameters.
From the data of the Super-Kamiokande
experiment it follows that the flux of 
$^8$B neutrinos is
\begin{equation}
\Phi_{^8\mathrm{B}}
=
( 2.44 \pm 0.10 ) \times 10^6 \, \mathrm{cm}^{-2} \, \mathrm{s}^{-1}
\,.
\end{equation}
If we subtract the contribution of
$^8$B neutrinos   
from the event rate observed in the Homestake experiment,
we obtain an upper bound for the
contribution of $^7$Be neutrinos to the chlorine event rate:
\begin{equation}
S^{^7\mathrm{B}}_{\mathrm{Cl}}
\leqslant
S^{\mathrm{exp}}_{\mathrm{Cl}} - S^{^8\mathrm{B}}_{\mathrm{Cl}}
=
-0.22 \pm 0.35 \ \mathrm{SNU}
\,.
\end{equation}
According to the SSM,
the contribution of  $^7$Be neutrinos to the Chlorine
event rate is significantly larger
($S^{^7\mathrm{B}}_{\mathrm{Cl}}(\mathrm{SSM}) = 1.1 \pm 0.1 \, \mathrm{SNU}$).
Such a strong
suppression of the flux of $^7$Be neutrinos cannot be explained by any
known astrophysical mechanism \cite{Bahcall-nu98}.
One can reach the same conclusion using the Super-Kamiokande and GALLEX-SAGE data.

The solar neutrinos problem can be solved if we assume that the solar
$\nu_e$'s transform into another flavor
($\nu_\mu$, $\nu_\tau$) or sterile states
through neutrino oscillations.
In order to explain all the existing solar neutrino data, it is sufficient to
assume that
transitions between two neutrino states alone take place.

The solar neutrinos produced in the thermonuclear reactions
in the central zone of the Sun 
pass through a large amount of matter on the way to the Earth.
If the value of the
parameter $\Delta{m}^2$
lies between
$\sim 10^{-8} \, \mathrm{eV}^2 $
and
$\sim 10^{-4} \, \mathrm{eV}^2 $,
coherent matter effects become important and
the transition probability of solar
$\nu_e$'s into other states can be resonantly enhanced
(MSW effect \cite{MSW}),
even for small values of the mixing angle $\vartheta$.
Assuming the validity of the SSM and that the MSW effects does indeed take place, 
the analysis of all the solar neutrino data leads to
the following two possible sets of
best-fit values of the oscillation parameters
\cite{Bahcall-Krastev-Smirnov-98,SK-sun-analysis-99,Nakahata-TAUP-99}
\begin{eqnarray}
&&
\Delta{m}^2 \simeq 5 \times 10^{-6} \, \mathrm{eV}^2
\,,
\qquad
\sin^2{2\vartheta} \simeq 5 \times 10^{-3}
\,,
\nonumber
\\
&&
\Delta{m}^2 \simeq 3.2 \times 10^{-5} \, \mathrm{eV}^2
\,,
\qquad
\sin^2{2\vartheta} \simeq 0.8
\,.
\end{eqnarray}
In the case of $\nu_e \to \nu_s$
transitions only small values of the mixing angle are allowed
\cite{Bahcall-Krastev-Smirnov-98,SK-sun-analysis-99,Concha-99,Nakahata-TAUP-99}.

The existing data can also be explained if 
$\Delta{m}^2 \ll 10^{-8} \, \mathrm{eV}^2$
and matter effects are unimportant.
In this case,
the $\nu_e$ survival probability
is given by 
the standard two neutrino vacuum expression (see Eq.(\ref{surv1})).
From the analysis of the data, the following best-fit values of the oscillation 
parameters have been found
\cite{Bahcall-Krastev-Smirnov-98,Barger-Whisnant-VO-99,Nakahata-TAUP-99}
\begin{equation}
\Delta{m}^2 \simeq 4.3 \times 10^{-10} \, \mathrm{eV}^2
\,,
\qquad
\sin^2{2\vartheta} \simeq 0.79
\,.
\end{equation}
This solution of the solar neutrino problem
is called ``vacuum oscillation solution''
or
``just-so vacuum solution''.

Recently
a new solar neutrino experiment, SNO \cite{SNO}, started in Canada.
In this experiment, the solar $^8$B neutrinos will be detected through the
charged current (CC)  process,
$\nu_e + d \to e^- +p +p$,
the neutral current (NC)  process, $\nu + d \to 
\nu + n + p$,
and elastic neutrino-electron scattering
$\nu +e \to \nu +e$.

From the detection of the solar neutrinos via the CC process, 
the spectrum
of $\nu_e$'s on the Earth can be measured and compared with
the well-known
$^8$B neutrino spectrum 
predicted by the theory of weak interactions.
A comparison of the measured spectrum with the predicted one can provide us with
a model-independent check on whether the solar 
$\nu_e$'s
have transformed into other states in a energy-dependent
way\footnote{The investigation
of a possible distortion of the
$^8$B neutrino spectrum is under investigation also in the Super-Kamiokande
experiment \cite{SK-sun,Nakahata-TAUP-99}.}.
The detection of neutrinos 
via the NC process
can determine the total flux of all
flavors, $\nu_e$, $\nu_\mu $ and  $\nu_\tau $. The comparison
of the NC and CC
measurements, 
will provide us with another possibility to check in a
model-independent way whether
transitions of the solar $\nu_e$'s into other flavor states actually take place. 

From all the analyses of the solar neutrino data it follows
that the flux of medium energy $^7$Be neutrinos is strongly suppressed.
A future solar neutrino experiment,
Borexino \cite{Borexino},
in which mainly $^7$Be neutrinos will be detected, can check
this general conclusion obtained from the existing data. 

Several other future solar neutrino experiments
\cite{Lanou-nu98}
(ICARUS,
GNO,
LENS,
HELLAZ
and others)
are under planning or development. In these experiments 
different parts of the solar
neutrino spectrum will be explored in detail.

A sensitivity to values of $\Delta{m}^2$
as small as $ \sim 10 ^{-5} \, \mathrm{eV}^2 $ will be reached
in the LBL reactor experiments Kam-Land in Japan
\cite{Kam-Land} and
Borexino in
Italy \cite{Borexino}
(with distances between reactors and detectors of about 200 km
and 800 km,
respectively).
These experiments will be able to check
the large mixing angle MSW
solution of the solar neutrino problem.

The third indication in favor of neutrino oscillations was reported by 
the Los Alamos neutrino
experiment LSND \cite{LSND-96-PRL,LSND-98-PRL}.
In this experiment, neutrinos were produced in
decays at rest of
$\pi^+$
($ \pi^+ \to \mu^+ \nu_\mu $)
and
$\mu^+$ ($ \mu^+ \to e^+ \nu_e \bar{\nu}_\mu $).
Thus, there are no $\bar\nu_e$'s from the source.
At a distance of about 30 m from the source 
$\bar\nu_e$'s are searched for  
in the large LSND detector
(about 180 tons of liquid scintillator)
via the process
\begin{equation}
\bar{\nu}_e + p \to e^+ + n
\,.
\end{equation}
A significant number of
$\bar\nu_e$ events
(22 events with an expected background of $ 4.6 \pm 0.6 $ events)
has been found
in the range of neutrino energy 
$ 36 \, \mathrm{MeV} < E < 60 \, \mathrm{MeV} $
\cite{LSND-96-PRL}.

The observed events can be explained by
$\bar\nu_\mu \to \bar\nu_e $ oscillations.
Taking into account the results of 
other reactor and accelerator experiments,
in which neutrino oscillations were not observed,
the following ranges for the oscillation parameters are allowed:
\begin{equation}
0.2 \, \mathrm{eV}^2 \lesssim \Delta{m}^2 \lesssim 2 \, \mathrm{eV}^2
\,,
\qquad
3 \times 10^{-3} \lesssim \sin^2{2\vartheta} \lesssim 4 \times 10^{-2}
\,.
\label{LSNDrange}
\end{equation}

The LSND Collaboration found also some evidence
in favor of $\nu_\mu\to\nu_e$ transitions of the $\nu_\mu$'s
generated by $\pi^+$ decay in flight
\cite{LSND-98-PRL}.
The $\nu_e$'s produced in this way can be distinguished
from the $\nu_e$'s generated by $\mu^+$ decay at rest
because they have higher energies.
The resulting allowed values of the parameters
$\Delta{m}^2$ and $\sin^22\vartheta$
are compatible with those in Eq.(\ref{LSNDrange}).

In another accelerator experiment, KARMEN
\cite{KARMEN}, designed to find
$\bar\nu_\mu \to \bar\nu_e $ oscillations, no positive signal was
found.
The sensitivity of the KARMEN experiment 
to the value of the parameter $\sin^2{2\vartheta}$
is, however, smaller than the
sensitivity of the LSND
experiment and at the moment there is no contradiction between the results
of these two experiments,
although part of the LSND allowed region
is disfavored by the results of the KARMEN experiment.
New experiments are needed in order to investigate in detail
the LSND anomaly.
Four such experiments have been proposed and are under study:
BooNE \cite{BooNE} at Fermilab,
I-216 \cite{I-216} at CERN,
ORLaND \cite{ORLaND} at Oak Ridge
and
NESS at the European Spallation Source \cite{NESS}.
Among these proposals BooNE is approved and will start in the year 2001.

In summary,
indications
in favor of nonzero neutrino masses and neutrino mixing have been found
in the atmospheric, solar and LSND neutrino experiments.
What conclusions can we drawn about
the possible neutrino mass spectra and
the values of the elements of the
neutrino mixing matrix $U$ from the data of these experiments?

Assuming the validity of the solar and atmospheric
indications in favor of neutrino oscillations,
the types of neutrino mass spectra allowed by the data
crucially depend on the validity of the LSND result.
If this result fails to be confirmed by future experiments,
three massive neutrinos
with the hierarchical mass spectrum of see-saw
type and with $\Delta{m}^2_{21}$
relevant for the oscillations of solar neutrinos
and $\Delta{m}^2_{31}$ relevant for the oscillations of atmospheric
neutrinos are enough to describe all the existing data
\cite{Fogli-sun-atm}.

The CHOOZ and Super-Kamiokande results suggest 
that, in the three neutrino case, 
the element $|U_{e3}|^2$ is small \cite{Vissani97,BG98-dec}.
This means that the oscillations of solar and atmospheric neutrinos are
practically decoupled and are described by the two-neutrino 
formalism.
Neutrinos are very light in this scenario:
the heaviest mass is
$m_3 \simeq 5 \times 10^{-2} \mathrm{eV}^2$.
In order to answer the fundamental question as to whether 
massive neutrinos are Dirac or Majorana particles, it is necessary to
increase the sensitivity of the experiments searching for neutrinoless
double-$\beta$ decay by at least one order of magnitude
\cite{BGKM98-bb,BGGKP-bb-99}.

\begin{figure}[t]
\begin{center}
\includegraphics[bb=98 644 513 773,width=0.99\linewidth]{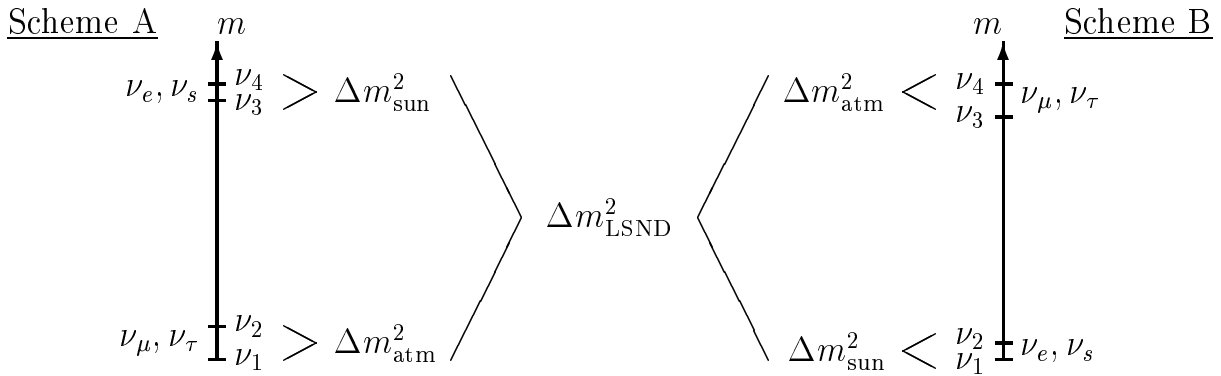}
\end{center}
\caption{ \label{schemesAB}
The two types of neutrino mass spectra that can accommodate 
the solar, atmospheric and LSND scales of $\Delta{m}^2$ and the
mixing schemes that emerge if the Big-Bang Nucleosynthesis
constraint on the number of light neutrinos is less than 4.}
\end{figure}

If the LSND result is confirmed,
at least
three different scales of 
$\Delta{m}^2$
(LSND, atmospheric and solar)
are needed in order to describe the data,
which implies that there are at least four massive (but light)
neutrinos.
In the minimal scheme with four massive neutrinos,
only the two mass spectra
\begin{equation}\label{spectrum}
\mbox{(A)}
\qquad
\underbrace{
\overbrace{m_1 < m_2}^{\mathrm{atm}}
\ll
\overbrace{m_3 < m_4}^{\mathrm{sun}}
}_{\mathrm{LSND}}
\qquad \mbox{and} \qquad
\mbox{(B)}
\qquad
\underbrace{
\overbrace{m_1 < m_2}^{\mathrm{sun}}
\ll
\overbrace{m_3 < m_4}^{\mathrm{atm}}
}_{\mathrm{LSND}}
\,,
\label{AB}
\end{equation}
with two couples of close masses separated by the
the "LSND gap" of about 1 eV, 
are compatible with all the existing data \cite{BGG-AB}.
The existence of four massive light neutrinos implies that
a sterile neutrino should exist
in addition to the flavor neutrinos
$\nu_e$, $\nu_\mu$ and $\nu_\tau$.
Furthermore,
if the standard Big-Bang Nucleosynthesis constraint
on the number of light neutrinos
\cite{BBN-standard}
is less than 4
\cite{art:Burles:1999zt},
there is a stringent limit on the mixing of the sterile neutrino
with the two massive neutrinos
that are responsible for the oscillations
of atmospheric neutrinos
and
the two allowed schemes have the form shown in Fig.~\ref{schemesAB}
\cite{Okada-Yasuda97,BGGS98-BBN},
\textit{i.e.}
$\nu_s$ is mainly mixed with the
two massive neutrinos that contribute to solar neutrino oscillations
($\nu_3$ and $\nu_4$ in scheme A
and
$\nu_1$ and $\nu_2$ in scheme B)
and
$\nu_\tau$ is mainly mixed with the
two massive neutrinos that contribute to the oscillations
of atmospheric neutrinos.

In the extended SM with massive neutrinos,
there is no room for sterile neutrinos.
Thus,
a successful explanation of \textit{all} the existing data requires new exciting 
physics.

\section{Conclusions}
\label{Conclusions}

The evidence in favor of oscillations of atmospheric neutrinos 
found in the Super-Kamiokande experiment and
the indications in favor of
oscillations obtained 
in other atmospheric neutrino experiments,
in the solar neutrino experiments
and in the LSND experiment
has opened a new chapter in neutrino
physics: \textit{the physics of massive and mixed neutrinos}.

There are many open problems in the physics of massive neutrinos.

We are now anxiously waiting for the results
of the new neutrino oscillation experiments SNO and K2K,
that started their operation in 1999, and
the future experiments Borexino, ICARUS,
BooNe, MINOS and many others,
that will start after the year 2001.
We hope that the results of these experiments
will answer the questions that have
been puzzling us for the past decades.

There is no doubt that the program of future investigations
of neutrino
oscillations will lead to a significant progress in understanding 
\textit{the origin of the tiny neutrino masses and of neutrino mixing},
which is undoubtedly of extreme importance 
for the future of 
elementary particle physics and astrophysics.

\end{document}